\documentclass[aps,twocolumn,superscriptaddress]{revtex4}
\usepackage{graphicx}

\begin{document}

\title{Interqubit coupling mediated by a high-excitation-energy
quantum object}

\

\author{S. Ashhab}
\affiliation{Frontier Research System, The Institute of Physical
and Chemical Research (RIKEN), Wako-shi, Saitama 351-0198, Japan}
\affiliation{Physics Department and Michigan Center for
Theoretical Physics, The University of Michigan, Ann Arbor,
Michigan 48109-1040, USA}

\author{A. O. Niskanen}
\affiliation{CREST-JST, Kawaguchi, Saitama 332-0012,Japan}
\affiliation{VTT Technical Research Centre of Finland, Sensors, PO
BOX 1000, 02044 VTT, Finland}

\author{K. Harrabi}
\affiliation{CREST-JST, Kawaguchi, Saitama 332-0012,Japan}

\author{Y. Nakamura}
\affiliation{Frontier Research System, The Institute of Physical
and Chemical Research (RIKEN), Wako-shi, Saitama 351-0198, Japan}
\affiliation{CREST-JST, Kawaguchi, Saitama 332-0012,Japan}
\affiliation{NEC Nano Electronics Research Laboratories, Tsukuba,
Ibaraki 305-8501, Japan}

\author{T. Picot}
\affiliation{Quantum Transport Group, Kavli Institute of
NanoScience, Delft University of Technology, Lorentzweg 1, 2628 CJ
Delft, The Netherlands}

\author{P. C. de Groot}
\affiliation{Quantum Transport Group, Kavli Institute of
NanoScience, Delft University of Technology, Lorentzweg 1, 2628 CJ
Delft, The Netherlands}

\author{C. J. P. M. Harmans}
\affiliation{Quantum Transport Group, Kavli Institute of
NanoScience, Delft University of Technology, Lorentzweg 1, 2628 CJ
Delft, The Netherlands}

\author{J. E. Mooij}
\affiliation{Quantum Transport Group, Kavli Institute of
NanoScience, Delft University of Technology, Lorentzweg 1, 2628 CJ
Delft, The Netherlands}

\author{Franco Nori}
\affiliation{Frontier Research System, The Institute of Physical
and Chemical Research (RIKEN), Wako-shi, Saitama 351-0198, Japan}
\affiliation{Physics Department and Michigan Center for
Theoretical Physics, The University of Michigan, Ann Arbor,
Michigan 48109-1040, USA}

\date{\today}

% PACS:
% 03.67.Lx    Quantum computation
% 85.25.Cp    Josephson devices
% 85.25.Hv    Superconducting logic elements and memory devices;
%             microelectronic circuits
% 03.67.Mn    Entanglement production, characterization, and manipulation
% 74.50.+r    Tunneling phenomena; point contacts, weak links,
%             Josephson effects
% 85.25.Am    Superconducting device characterization,
%             design, and modeling

\begin{abstract}

We consider a system composed of two qubits and a
high-excitation-energy quantum object used to mediate coupling
between the qubits. We treat the entire system quantum
mechanically and analyze the properties of the eigenvalues and
eigenstates of the total Hamiltonian. After reproducing well-known
results concerning the leading term in the mediated coupling, we
obtain an expression for the residual coupling between the qubits
in the off state. We also analyze the entanglement between the
three objects, i.e.~the two qubits and the coupler, in the
eigenstates of the total Hamiltonian. Although we focus on the
application of our results to the recently realized
parametric-coupling scheme with two qubits, we also discuss
extensions of our results to harmonic-oscillator couplers,
couplers that are near resonance with the qubits and multi-qubit
systems. In particular, we find that certain errors that are
absent for a two-qubit system arise when dealing with multi-qubit
systems.

\end{abstract}

\maketitle

\section{Introduction}

Superconducting qubits are among the main candidates for the
possible implementation of quantum-information-processing tasks
\cite{YouReview}. Coherent dynamics of a single qubit has been
achieved at various laboratories. Several interesting two-qubit
experiments have also been performed
\cite{Pashkin,Johnson,Berkley,Yamamoto,Izmalkov,Xu,McDermott,Majer1,PlourdeExp,GrajcarExp,Steffen,vdPloeg,Harris,Hime,NiskanenExp1,NiskanenExp2,Plantenberg,Sillanpaa,Majer2}.
The early experiments
\cite{Pashkin,Johnson,Berkley,Yamamoto,Izmalkov,Xu,McDermott,Majer1,PlourdeExp,GrajcarExp,Steffen}
were limited to fixed inter-qubit coupling. In order to scale up
qubit circuits, however, it is highly desirable to be able to tune
{\it in situ} the coupling strengths between the different qubits.
The idea of coupling two qubits to a high-excitation-energy
quantum object that would mediate coupling between the qubits
\cite{Averin,PlourdeTh,Wallquist1,MaassenvdBrink,Makhlin,Kim} has
lead to experimental demonstrations of tunable coupling
\cite{vdPloeg,Harris,Hime}. However, Refs.~\cite{vdPloeg,Harris}
probed the magnetic properties of the circuit in its ground state;
thus the approach used there is not suited to implement the
standard gate-based quantum computing, but possibly adiabatic
quantum computing \cite{Farhi}, where the quantum register is
ideally never excited. In Ref.~\cite{Hime} tunability of the
coupling was demonstrated through spectroscopic measurements with
the qubits biased away from their coherence optimal points.
Generalizing the idea of mediated coupling
\cite{Averin,PlourdeTh,Wallquist1,MaassenvdBrink,Makhlin,Kim} to
parametric tunable coupling
\cite{Bertet,NiskanenTh,GrajcarTh,Liu}, it was proposed that one
can bias the qubits at their optimal points and also adjust the dc
component of the mediated coupling to cancel the direct
inter-qubit coupling throughout the experiment. Applying a
microwave pulse to the coupler at the sum or difference frequency
of the qubits' characteristic frequencies would then turn on the
coupling, but only for the duration of the applied microwave
pulse. This proposal, combining tunable coupling and long
coherence times, was realized experimentally in
Ref.~\cite{NiskanenExp2}.

Another related direction of growing research activity is the idea
of using a harmonic-oscillator `cavity' as a data bus with the
potential that a single cavity could mediate coupling between a
large number of qubits
\cite{Sillanpaa,Majer2,Chiorescu,CavityTheory}. If the harmonic
oscillator mediates coupling via high-energy virtual excitations,
it can be described similarly to other high-excitation-energy
couplers. Couplers that are near resonance with the qubits, but
sufficiently detuned such that they mediate coupling through
virtual excitations, can also be treated using a similar approach.
We shall show, however, that it would be rather difficult to
achieve tunability in the coupling in the case of
harmonic-oscillator couplers.

Early theoretical studies on tunable couplers have generally taken
the semiclassical approach, which we shall explain below. The
semiclassical treatment is sufficient to evaluate the leading term
in the effective coupling mediated by the coupler. As qubit
circuits that include couplers are becoming an experimental
reality, however, there is an increasing need for a more careful
analysis of these quantum couplers. Some recent studies
\cite{NiskanenTh,Hutter} used a number of quantum-mechanical
techniques and obtained results beyond the semiclassical
calculations. For example, Ref.~\cite{Hutter} derived an
expression for the residual coupling term when the main coupling
channel is turned off. Here we take a fully quantum approach,
where we analyze the properties of the eigenvalues and eigenstates
of the total Hamiltonian. We obtain results that were not captured
by previous studies, particularly results concerning the ideality
of the off state in this coupling scheme.

The main questions of interest that the quantum treatment can be
used to answer include:
(1) Can we make the coupler's energy splitting very large but
still maintain the mediated coupling between the qubits?
(2) Can we turn the coupling off completely? In other words, will
there be any residual coupling terms if the system is biased such
that the main coupling channel is at its zero point?
(3) How much entanglement is there between the qubits and the
coupler, and how does this entanglement affect a realistic
experimental setup?
These questions will be answered in the analysis below.

The paper is organized as follows. In Sec.~II we describe the
system and its Hamiltonian. In Sec.~III we review the
semiclassical approach and discuss what predictions we can expect
from it regarding the ideality of the off state. In Sec.~IV we
perform the fully quantum analysis of the problem: We derive
expressions for the leading-order effective coupling strength
mediated by the coupler, the residual coupling in the off state
and the amount of entanglement between the qubits and the coupler;
We also discuss the extension of our results to the case of a
harmonic-oscillator coupler and that of a coupler that is almost
resonant with the qubits. We discuss in Sec.~V the implications of
the results obtained in Sec.~IV with regard to present-day and
future experiments. In this context we also consider multi-qubit
systems. Section VI contains concluding remarks. Some details of
the calculations are explained in the Appendix.

\section{System and Hamiltonian}

Let us take a system composed of two qubits and a third object
that we would like to use as a coupler. Although tunability is
considered the main advantage of this coupling scheme, for the
purpose of answering the questions of main interest to us it
suffices to focus on the case where the external bias parameters
are set to fixed values. We therefore treat a time-independent
Hamiltonian. Since we are assuming that we can speak of three
distinct quantum objects, we must be able to write down a
Hamiltonian that reflects this clear separation of the different
objects in the system. We therefore express the Hamiltonian as:
\begin{equation}
\hat{H} = \hat{H}_1 + \hat{H}_2 + \hat{H}_{\rm C} + \hat{H}_{12} +
\hat{H}_{\rm 1C} + \hat{H}_{\rm 2C},
\label{eq:Hamiltonian}
\end{equation}
where the first three terms are the Hamiltonians of the separate
objects in the system, and the last three terms describe coupling
between those objects. As a realistic, representative case, we
take the different terms in the Hamiltonian to have the forms:
\begin{eqnarray}
\hat{H}_1 & = & \frac{\Delta_1}{2} \hat{\sigma}_z^{(1)} +
\frac{\epsilon_1}{2} \hat{\sigma}_x^{(1)}
\\
\hat{H}_2 & = & \frac{\Delta_2}{2} \hat{\sigma}_z^{(2)} +
\frac{\epsilon_2}{2} \hat{\sigma}_x^{(2)}
\\
\hat{H}_{\rm C} & = & \left(
\begin{array}{cccc}
0 & 0 & 0 & \cdots \\
0 & \eta_1 & 0 & \\
0 & 0 & \eta_2 & \\
\vdots & & & \ddots \\
\end{array}
\right)
\\
\hat{H}_{12} & = & J_0 \; \hat{\sigma}_x^{(1)} \otimes
\hat{\sigma}_x^{(2)} \nonumber
\\
& = & \left(
\begin{array}{cccc}
0 & 0 & 0 & J_0 \\
0 & 0 & J_0 & 0 \\
0 & J_0 & 0 & 0 \\
J_0 & 0 & 0 & 0 \\
\end{array}
\right)
\\
\hat{H}_{\rm 1C} & = & \hat{\sigma}_x^{(1)} \otimes \hat{A}
\\
\hat{H}_{\rm 2C} & = & \hat{\sigma}_x^{(2)} \otimes \hat{B},
\end{eqnarray}
with the coupler energies $0,\eta_1,\eta_2,...$ arranged in
increasing order,
\begin{equation}
\hat{A} = \left(
\begin{array}{cccc}
A_{00} & A_{01} & A_{02} & \cdots \\
A_{10} & A_{11} & A_{12} & \\
A_{20} & A_{21} & A_{22} & \\
\vdots & & & \ddots \\
\end{array}
\right),
\end{equation}
and similarly for $\hat{B}$. For the case of superconducting flux
qubits, $\Delta_j$ is the minimum gap of qubit $j$, $\epsilon_j$
represents the deviation from the degeneracy point of half-integer
flux quantum threading the qubit loop [i.e.~$\epsilon_j=I_{p,j}
(\Phi_{\rm ext, \it j}-\Phi_0/2)$, where $I_{p,j}$ is the
persistent current of qubit $j$, $\Phi_{\rm ext, \it j}$ is the
externally applied flux in the loop of qubit $j$, and $\Phi_0$ is
the flux quantum], $J_0$ is the direct qubit-qubit coupling
strength, and $\hat{\sigma}_{\alpha}^{(j)}$ are the usual Pauli
matrices of qubit $j$. Note that the minimum gap $\Delta_j$ is the
coefficient of $\hat{\sigma}_z^{(j)}$ above, in contrast with some
alternative conventions used in the literature where $\Delta_j$ is
the coefficient of $\hat{\sigma}_x^{(j)}$. Note also that the
operators $\hat{A}$ and $\hat{B}$ must satisfy the relations
$A_{ij}=A_{ji}^*$ and $B_{ij}=B_{ji}^*$. The coupler's Hamiltonian
$\hat{H}_{\rm C}$ is written in its own eigenbasis (thus it is
diagonal), and its ground-state energy has been set to zero. We
shall express Hamiltonians and energies in frequency units
throughout this paper.

We now make the assumption that the largest energy scale in the
problem is the excitation energy of the coupler. In other words
$\eta_1$ is larger than any relevant energy scale in the
Hamiltonian excluding $\hat{H}_{\rm C}$. We shall not make any
assumption regarding the relation between the qubit energy scale
$\sqrt{\Delta_j^2+\epsilon_j^2}$ and the qubit-coupler coupling
energy scale (i.e.~the scale of the matrix elements in $\hat{A}$
and $\hat{B}$). Although it is not crucial for most of our
analysis below, we shall generally take the direct coupling energy
scale $J_0$ to be smaller than the qubit energy scale.

\section{Semiclassical treatment}

Before embarking on the semiclassical description of our system,
it is instructive to recall the problem of calculating interatomic
forces within the hydrogen molecule \cite{Baym}. One starts by
assuming that the nuclei have fixed locations in space, leaving
the degrees of freedom associated with electron motion as the only
variables in the problem. The ground state of the electronic
degrees of freedom is calculated and expressed as a function of
the distance between the nuclei. At this point, the total energy
(direct Coulomb energy of the nuclei plus electronic ground-state
energy) is expressed as a function of the relative position
operator between the nuclei, and the nuclear motion is treated
quantum mechanically. One can now obtain information related to
atomic motion or molecular states without having to worry about
electronic motion. The first step in the calculation ensures that
the effects of the electrons' adiabatic adjustment to nuclear
motion are properly taken into account. The reason why the
electronic degrees of freedom can adjust adiabatically to nuclear
motion is that they are associated with a much higher energy
scale, and thus changes that result from nuclear motion are felt
by the electrons as very slow variations.

A similar procedure can be applied when dealing with inter-qubit
couplers. Since the qubit-coupler interactions contain the
operators $\hat{\sigma}_x^{(j)}$, one first assumes that the
qubits are in eigenstates of $\hat{\sigma}_x^{(j)}$. One therefore
needs to calculate the ground state energy of the Hamiltonian (or,
more precisely, the four Hamiltonians):
\begin{eqnarray}
\hat{H}_{\rm C,eff} & \equiv & \hat{H}_{\rm C} + \hat{H}_{\rm 1C}
+ \hat{H}_{\rm 2C} \nonumber
\\
& \stackrel{\sigma_x^{(j)}=\pm 1}{\longrightarrow} & \hat{H}_{\rm
C} \pm \hat{A} \pm \hat{B}.
\label{eq:CouplerAdiabHam}
\end{eqnarray}
The ground state energy [i.e.~the four values obtained from
Eq.~(\ref{eq:CouplerAdiabHam})] can then be expressed in the form
\cite{NoCJ}:
\begin{equation}
E_0 \left\{ \hat{H}_{\rm C,eff} \right\} = c_1 + c_2
\sigma_x^{(1)} + c_3 \sigma_x^{(2)} + c_4 \sigma_x^{(1)} \otimes
\sigma_x^{(2)}.
\end{equation}
Note that we have started with the assumption that the qubits have
well-defined values of $\sigma_x^{(j)}$. One could therefore say
that the above expression is not a quantum operator. However, it
is straightforward to follow the adiabaticity argument: the
coupler will, to a very good approximation, always be in the
ground state that corresponds to the instantaneous values of
$\sigma_x^{(j)}$ (including the possibility of quantum
superpositions). One can therefore turn back to the qubits and
analyze their dynamics using the effective Hamiltonian:
\begin{equation}
\hat{H}_{\rm q,eff} = \hat{H}_1 + \hat{H}_2 + \hat{H}_{12} +
\hat{H}_{\rm mediated},
\end{equation}
where
\begin{equation}
\hat{H}_{\rm mediated} = c_1 + c_2 \hat{\sigma}_x^{(1)} + c_3
\hat{\sigma}_x^{(2)} + c_4 \hat{\sigma}_x^{(1)} \otimes
\hat{\sigma}_x^{(2)}.
\label{eq:Hmediated}
\end{equation}
The above derivation shows that the response of the coupler to
external perturbations (induced by the qubits and represented by
the terms $\pm \hat{A}$ and $\pm \hat{B}$) translates into a
renormalization of the qubit bias points [second and third term in
Eq.~(\ref{eq:Hmediated})] and an additional coupling term between
the qubits [last term in Eq.~(\ref{eq:Hmediated})].

It is interesting to note here that in the semiclassical treatment
above, the effect of the coupler on the qubits is completely
described by the Hamiltonian $\hat{H}_{\rm mediated}$. Therefore,
if the bias point is chosen such that the coefficient $c_4$ in
Eq.~(\ref{eq:Hmediated}) cancels the direct coupling strength
$J_0$, there would be no residual coupling between the qubits. The
off state would therefore correspond to complete decoupling
between the qubits. In contrast, we shall show below that the
fully quantum treatment predicts the presence of finite residual
coupling effects.

\section{Quantum treatment}

We now take the Hamiltonian of Sec.~II (Eq.~\ref{eq:Hamiltonian})
and treat it quantum mechanically. In particular, we would like to
analyze the properties of the eigenvalues and eigenstates of the
total Hamiltonian. For the purposes of the calculations in this
section, we divide the Hamiltonian into two parts as follows:
\begin{equation}
\hat{H} = \hat{H}_0 + \hat{V},
\end{equation}
where
\begin{eqnarray}
\hat{H}_0 & = & \frac{\Delta_1}{2} \hat{\sigma}_z^{(1)} +
\frac{\Delta_2}{2} \hat{\sigma}_z^{(2)} + \hat{H}_{\rm C}
\\
\hat{V} & = & \frac{\epsilon_1}{2} \hat{\sigma}_x^{(1)} +
\frac{\epsilon_2}{2} \hat{\sigma}_x^{(2)} + \hat{H}_{12} +
\hat{H}_{\rm 1C} + \hat{H}_{\rm 2C}.
\end{eqnarray}
We use the basis $\{ |00\underline{0}\rangle,
|01\underline{0}\rangle, |10\underline{0}\rangle,
|11\underline{0}\rangle, |00\underline{1}\rangle,
|01\underline{1}\rangle$, $|10\underline{1}\rangle,
|11\underline{1}\rangle, |00\underline{2}\rangle, ... \}$, where
the first, second and third quantum numbers describe,
respectively, the state of the first qubit, second qubit and
coupler (the state of the coupler will be distinguished from those
of the qubits using an underline throughout this paper); note that
we allow the coupler to have more than two levels, and we shall
assume an infinite number of levels in the expressions below. We
can now express $\hat{H}_0$ as
\begin{widetext}
\begin{equation}
\hat{H}_0 = \left(
\begin{array}{ccccccccc}
\frac{-\Delta_1-\Delta_2}{2} & 0 & 0 & 0 & & & & & \cdots \\
0 & \frac{-\Delta_1+\Delta_2}{2} & 0 & 0 & & 0 & & & \\
0 & 0 & \frac{\Delta_1-\Delta_2}{2} & 0  & & & & & \\
0 & 0 & 0 & \frac{\Delta_1+\Delta_2}{2}  & & & & & \\
& & & & \eta_1+\frac{-\Delta_1-\Delta_2}{2} & 0 & 0 & 0 & \\
& 0 & & & 0 & \eta_1+\frac{-\Delta_1+\Delta_2}{2} & 0 & 0 & \\
& & & & 0 & 0 & \eta_1+\frac{\Delta_1-\Delta_2}{2} & 0  & \\
& & & & 0 & 0 & 0 & \eta_1+\frac{\Delta_1+\Delta_2}{2}  & \\
\vdots & & & & & & & & \ddots\\
\end{array}
\right).
\label{eq:BigH0}
\end{equation}
\end{widetext}

\subsection{Coupling strength: leading term}

In order to calculate the effective coupling strength, we now
calculate the energies of the lowest four energy levels. We
therefore want to construct a $4 \times 4$ effective-Hamiltonian
matrix describing the lowest energy levels while taking into
account the effects of the higher levels.

We now follow a standard calculation \cite{Sakurai} (see Appendix)
that gives:
\begin{equation}
\hat{H}_{\rm eff} \approx \hat{H}_{\rm eff}^{(0)} + \hat{H}_{\rm
eff}^{(1)} + \hat{H}_{\rm eff}^{(2)}
\end{equation}
where
\begin{widetext}
\begin{equation}
\hat{H}_{\rm eff}^{(0)} = \hat{H}_1 + \hat{H}_2 + \hat{H}_{12} =
\frac{1}{2} \left(
\begin{array}{cccc}
-\Delta_1-\Delta_2 & \epsilon_2 & \epsilon_1 & 2 J_0 \\ [8pt]
\epsilon_2 & -\Delta_1+\Delta_2 & 2 J_0 & \epsilon_1 \\ [8pt]
\epsilon_1 & 2 J_0 & \Delta_1-\Delta_2 & \epsilon_2 \\  [8pt]
2 J_0 & \epsilon_1 & \epsilon_2 & \Delta_1+\Delta_2 \\
\end{array}
\right) \nonumber
\end{equation}
\begin{equation}
\hat{H}_{\rm eff}^{(1)} = A_{00} \sigma_x^{(1)} + B_{00}
\sigma_x^{(2)} = \left(
\begin{array}{cccc}
0 & B_{00} & A_{00} & 0 \\
B_{00} & 0 & 0 & A_{00} \\
A_{00} & 0 & 0 & B_{00} \\
0 & A_{00} & B_{00} & 0 \\
\end{array}
\right) \nonumber
\end{equation}
\begin{equation}
\hat{H}_{\rm eff}^{(2)} = \sum_{l=1}^{\infty} \sum_{k=00,01,10,11}
\hat{P} \frac{ \hat{V} | k,l \rangle \langle k,l | \hat{V}
}{\hat{H}_{\rm eff}-E_{k,l}} \hat{P},
\label{eq:2ndOrderCorrection}
\end{equation}
\end{widetext}
where the operator $\hat{P}$ projects the state onto the space of
the lowest four eigenstates of $H_0$ ($|00\underline{0}\rangle,
|01\underline{0}\rangle, |10\underline{0}\rangle$ and
$|11\underline{0}\rangle)$; alternatively, one could say that the
operator $\hat{P}$ removes the size mismatch between the
four-dimensional Hilbert space of interest and the
infinite-dimensional operator $\hat{V}$. The sum over $l$ in
Eq.~(\ref{eq:2ndOrderCorrection}) runs over states where the
coupler is in one of its excited states. The energies $E_{k,l}$
are the eigenvalues of $\hat{H}_0$ with the qubits in state $k$
and the coupler in state $l$. In order to find the lowest-order
expression for the mediated coupling term, we make the
approximation $\hat{H}_{\rm eff}-E_{k,l} \approx - \eta_l$ in
Eq.~(\ref{eq:2ndOrderCorrection}). Thus we obtain the expression
\begin{eqnarray}
\hat{H}_{\rm eff}^{(2)} & = & \hat{H}_{\rm coupling}^{(2)} +
\hat{H}_{\rm shift}^{(2)}
\\
\hat{H}_{\rm coupling}^{(2)} & \approx & \left(
\begin{array}{cccc}
0 & 0 & 0 & -J_1 \\
0 & 0 & -J_1 & 0 \\
0 & -J_1 & 0 & 0 \\
-J_1 & 0 & 0 & 0 \\
\end{array}
\right) = - J_1 \hat{\sigma}_x^{(1)} \otimes \hat{\sigma}_x^{(2)}
\nonumber
\\
\hat{H}_{\rm shift}^{(2)} & \approx & - \Delta_{\rm shift} \left(
\begin{array}{cccc}
1 & 0 & 0 & 0 \\
0 & 1 & 0 & 0 \\
0 & 0 & 1 & 0 \\
0 & 0 & 0 & 1 \\
\end{array}
\right)
\nonumber
\end{eqnarray}
where
\begin{eqnarray}
J_1 & = & \sum_{l=1}^{\infty} \frac{A_{0l} B_{l0} + B_{0l}
A_{l0}}{\eta_l} \nonumber
\\
\Delta_{\rm shift} & = & \sum_{l=1}^{\infty} \frac{|A_{0l}|^2 +
|B_{0l}|^2}{\eta_l}.
\label{eq:J1DeltaShift}
\end{eqnarray}
The overall shift $\hat{H}_{\rm shift}^{(2)}$ does not have any
physical consequences to this order of the calculation and can be
neglected.

The effective Hamiltonian can now be expressed as
\begin{equation}
\hat{H}_{\rm eff} = \tilde{H}_0 + \tilde{H}_{\rm coupling},
\label{eq:EffectiveH}
\end{equation}
where
\begin{eqnarray}
\tilde{H}_0 & = & \frac{\Delta_1}{2} \hat{\sigma}_z^{(1)} +
\frac{\tilde{\epsilon}_1}{2} \hat{\sigma}_x^{(1)} +
\frac{\Delta_2}{2} \hat{\sigma}_z^{(2)} +
\frac{\tilde{\epsilon}_2}{2} \hat{\sigma}_x^{(2)} \nonumber
\\
\tilde{H}_{\rm coupling} & = & J \hat{\sigma}_x^{(1)} \otimes
\hat{\sigma}_x^{(2)}, \nonumber
\end{eqnarray}
with the parameters
\begin{eqnarray}
\tilde{\epsilon}_1 & = & \epsilon_1 + 2 A_{00} \nonumber
\\
\tilde{\epsilon}_2 & = & \epsilon_2 + 2 B_{00} \nonumber
\\
J & = & J_0 - J_1.
\end{eqnarray}
The above results agree with those of Ref.~\cite{NiskanenTh} when
the parameters of our model are taken to correspond to those
considered in Ref.~\cite{NiskanenTh}.

First we mention the physical interpretation of the terms
$2A_{00}$ and $2B_{00}$ in the expressions for
$\tilde{\epsilon}_j$. Taking the experimentally relevant case of
flux qubits as an example, these terms describe the fluxes
generated by the coupler (in its ground state) and going through
the qubit loops. As a result, if one wishes to bias the qubits at
their optimal points, the externally applied fluxes through the
qubit loops are not set to $\Phi_0/2$, but they are shifted from
that value to compensate for the coupler-induced contributions.
This is the physical explanation of the difference between
$\epsilon_j$ and $\tilde{\epsilon}_j$ \cite{A00B00}.

We can now answer the question of how high the coupler's energy
splitting can be. We note that the mediated coupling strength
$J_1$ is second order in the scale of $\hat{A}$ and $\hat{B}$, and
it is inversely proportional to the scale of $\eta_l$. Although
this means that we cannot simply take $\eta_1\rightarrow\infty$
keeping the other parameters fixed (otherwise $J_1\rightarrow 0$),
it also means that if we increase $A_{nm}$, $B_{nm}$ and $\eta_l$
(keeping the ratio $A_{0l}B_{l0}/\eta_l$ fixed) we can in fact
take $\eta_1\rightarrow\infty$ while maintaining the same level of
mediated coupling. We shall see below that this situation is
desirable for purposes of reducing residual-coupling effects.

We are also in a position to comment on the question of
monostability of the coupler (i.e.~the idea that a single energy
level of the coupler is relevant in the system under
consideration). The presence of the matrix elements $A_{l0}$ and
$B_{l0}$, which describe coupling between the coupler's ground and
excited states, in the expression for the mediated coupling
strength (Eq.~\ref{eq:J1DeltaShift}) demonstrates that the excited
states of the coupler play an important role in the coupling
mechanism. The semiclassical treatment relies on the fact that the
relevant information contained in these matrix elements is also
encoded, and more easily accessible experimentally, in the
response of the coupler's ground state to weak perturbations. As a
result, knowledge of the matrix elements themselves is not
necessary; knowledge of simple response parameters is sufficient
in order to calculate the mediated coupling strength.

\subsection{Residual coupling in the off state}

In Sec.~IV.A we have derived the leading-order terms in the effect
of the coupler on the two-qubit system. The results agree with the
predictions of the semiclassical treatment (although no explicit
expressions were given in Sec.~III). Our main motivation for using
the fully quantum treatment, however, is that it allows us to go
further in the calculation, e.g.~evaluating any off-state residual
coupling terms in the two-qubit effective Hamiltonian. Such closer
examination of the off state will be the subject of this and the
following subsection (as well as part of Sec.~V).

Since we are interested in an ideal off state with the qubits at
their optimal points, we consider the case where
$J=\tilde{\epsilon}_1=\tilde{\epsilon}_2=0$ (we shall allow $J$ to
take nonzero values in the course of the discussion below). In
order to enhance the robustness of the decoupling between the
qubits, we take $\Delta_1\neq\Delta_2$. The effective Hamiltonian
(Eq.~\ref{eq:EffectiveH}) is diagonal to lowest order in this
case. We can therefore proceed with calculating higher-order
corrections using the states $|00\underline{0}\rangle$,
$|01\underline{0}\rangle$, $|10\underline{0}\rangle$ and
$|11\underline{0}\rangle$.

The calculation of the residual-coupling Hamiltonian is now
carried out by calculating the energies of the lowest four levels
of the entire system (we shall refer to them as
$E_{00\underline{0}}$, $E_{01\underline{0}}$,
$E_{10\underline{0}}$, and $E_{11\underline{0}}$). The four values
that we obtain for the energies in the four-level spectrum can
then be used to extract four quantities: (1) an overall energy
that can be neglected, (2,3) the corrected (i.e.~renormalized)
values of the qubit splittings $\Delta_j$ and (4) a
residual-coupling energy (given by
$E_{00\underline{0}}+E_{11\underline{0}} -
E_{01\underline{0}}-E_{10\underline{0}}$). It is this last
quantity that is of most interest to us here. It characterizes a
coupling Hamiltonian of the form:
\begin{equation}
\hat{H}_{\rm residual} = J_{\rm residual} \; \hat{\sigma}_z^{(1)}
\otimes \hat{\sigma}_z^{(2)},
\label{eq:Hresidual}
\end{equation}
where
\begin{equation}
J_{\rm residual} = \frac{E_{00\underline{0}}+E_{11\underline{0}}
-E_{01\underline{0}}-E_{10\underline{0}}}{4}.
\end{equation}
Note that in writing Eq.~(\ref{eq:Hresidual}) we have used our
assumption that the qubits are biased at their optimal points.

Although one can make analytic progress calculating the energies
using the pseudo-degenerate perturbation theory approach, reaching
the relevant results requires several steps of the iterative
procedures explained in the Appendix (the lowest-order corrections
to the individual energies do not contribute to the combination
$E_{00\underline{0}}+E_{11\underline{0}}
-E_{01\underline{0}}-E_{10\underline{0}}$). We therefore only
present the results of numerical calculations that find the energy
levels of the entire system. It also turns out that proceeding
with the rather general model used so far complicates the
extraction of the important results. We therefore have to make
some simplifying assumptions. From now on, we focus on a simple
case that also happens to be relevant to recent experiments: We
take the coupler to be a two-level system, such that
\begin{eqnarray}
\hat{H}_{\rm C} & = & \frac{\epsilon_{\rm C}}{2}
\hat{\sigma}_x^{\rm (C)} + \frac{\Delta_{\rm C}}{2}
\hat{\sigma}_z^{\rm (C)} \nonumber
\\
& = & \frac{\Delta_{\rm C}}{2} \left( \tan\theta_{\rm C} \;
\hat{\sigma}_x^{\rm (C)} + \hat{\sigma}_z^{\rm (C)} \right)
\nonumber
\\
\hat{H}_{\rm 1C} & = & J_{\rm 1C} \; \hat{\sigma}_x^{(1)} \otimes
\sigma_x^{\rm (C)} \nonumber
\\
\hat{H}_{\rm 2C} & = & J_{\rm 2C} \; \hat{\sigma}_x^{(2)} \otimes
\sigma_x^{\rm (C)}
\end{eqnarray}
(note that, unlike the convention used in Sec.~II, $\hat{H}_{\rm
C}$ is not diagonal here). We therefore have
$\eta_1=\sqrt{\Delta_{\rm C}^2+\epsilon_{\rm C}^2}=\Delta_{\rm
C}/|\cos\theta_{\rm C}|$, $A_{00}=-A_{11}=-J_{\rm 1C}
\sin\theta_{\rm C}$, $A_{10}=A_{01}=J_{\rm 1C} \cos\theta_{\rm C}$
and similarly for $\hat{B}$. As a result, we find from
Eq.~(\ref{eq:J1DeltaShift}) that
\begin{equation}
J = J_0 - \frac{2 J_{\rm 1C} J_{\rm 2C} \cos^3\theta_{\rm
C}}{\Delta_{\rm C}}.
\label{eq:J}
\end{equation}

For clarity and definiteness in the following analysis, we now
focus on the parametric-coupling scheme
\cite{Bertet,NiskanenTh,GrajcarTh}, and we start by noting three
factors that can contribute to determining the ideal dc bias point
(it is worth mentioning at this point that in general it will not
be possible, nor necessary, to satisfy all three conditions
simultaneously). Firstly, we note from Eq.~(\ref{eq:J}) that the
range of values for the effective coupling strength $J$ extends
from $J_0-2J_{\rm 1C}J_{\rm 2C}/\Delta_{\rm C}$ (at $\theta_{\rm
C}=0$) to $J_0$ (at $\theta_{\rm C}=\pi/2$). In order to maximize
the effective coupling strength of the parametric-coupling scheme,
it would be desirable to set the dc bias point such that
$J=J_0-J_{\rm 1C}J_{\rm 2C}/\Delta_{\rm C}$, i.e.~halfway between
the two extremes; This situation would be obtained by setting
$\theta_{\rm C} \approx 0.65 \approx \pi/5$, irrespective of the
specific system parameters. At this bias point, the allowed
amplitude of the ac driving signal (i.e.~before encountering
nonlinearities in $J$) is maximized (see Fig.~1a). Secondly, it is
desirable to set $J=0$ at the dc bias point. Thirdly, it is also
desirable to set the residual coupling energy $J_{\rm
residual}=0$.

We now proceed with the numerical calculations as follows: We
first fix the parameters $\Delta_1$, $\Delta_2$, $\Delta_{\rm C}$
and $J_0$ (e.g.~$\Delta_1=4$ GHz, $\Delta_2=5$ GHz, $\Delta_{\rm
C}=30$ GHz and $J_0=50$ MHz). The coupling strengths $J_{\rm 1C}$
and $J_{\rm 2C}$ are then chosen as $J_{\rm 1C}=J_{\rm 2C}$ and
$J_{\rm 1C} J_{\rm 2C}/\Delta_{\rm C}=J_0$. This choice ensures
that a single dc bias point satisfies at least two of the three
desirable conditions mentioned above, i.e.~$J$ vanishes while
maximizing the achievable effective coupling strength for the
parametric-coupling scheme. The coupler's bias point, determined
by the angle $\theta_{\rm C}$, is now treated as the only variable
in the problem. For each value of $\theta_{\rm C}$, we
(numerically) bias the qubits at their optimal points (determined
by simultaneously minimizing both qubit energy splittings) and
calculate the residual coupling energy. The results are shown in
Fig.~1. We can clearly see that there is some residual coupling in
the off state (i.e.~at $\theta_{\rm C}\approx 0.65$); however, the
residual coupling strength in Fig.~1 is very small compared to
typical relaxation and dephasing times. After varying all the
parameters by at least a factor of 2 in either direction, we can
identify that the residual-coupling term is given by
\begin{widetext}
\begin{equation}
J_{\rm residual} \approx 4 \frac{\Delta_1 \Delta_2 J_0 J_{\rm 1C}
J_{\rm 2C}}{\Delta_{\rm C}^4} \cos^6\theta_{\rm C} - \alpha
\frac{\Delta_1 \Delta_2 J_{\rm 1C}^2 J_{\rm 2C}^2}{\Delta_{\rm
C}^5} \cos^5\theta_{\rm C} \sin^2(2\theta_{\rm C}),
\label{eq:ResidualJ}
\end{equation}
\end{widetext}
where the coefficient $\alpha$ is approximately 20.

\begin{figure}[h]
\includegraphics[width=8.0cm]{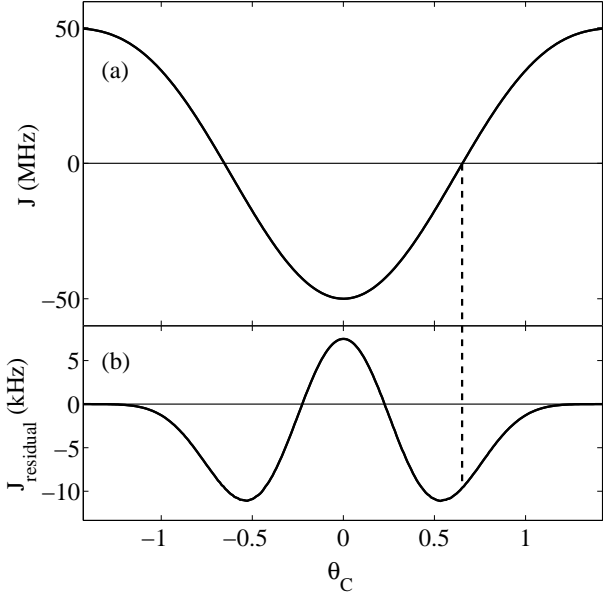}
\caption{Coefficients of the main (a) and residual (b) coupling
terms (i.e.~$J$ and $J_{\rm residual}$) as functions of the
coupler's bias point $\theta_{\rm C}$. In generating this figure
we took $\Delta_1=4$ GHz, $\Delta_2=5$ GHz, $\Delta_{\rm C}=30$
GHz, $J_0=50$ MHz and $J_{\rm 1C}=J_{\rm 2C}=\sqrt{J_0 \Delta_{\rm
C}}$. It should be noted, however, that the results are only
weakly dependent on the exact choice of parameters.}
\end{figure}

\begin{figure}[h]
\includegraphics[width=8.0cm]{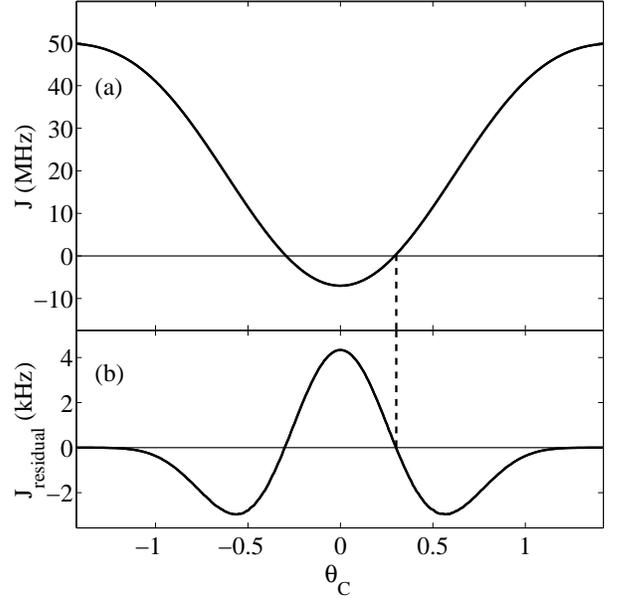}
\caption{Same as in Fig.~1, except that $J_{\rm 1C}=J_{\rm
2C}=0.75 \sqrt{J_0 \Delta_{\rm C}}$. As opposed to Fig.~1, $J$ and
$J_{\rm residual}$ now vanish at the same bias point.}
\end{figure}

It is encouraging for future experimental efforts that the
residual-coupling term obtained above decreases rapidly with
increasing coupler gap $\Delta_{\rm C}$, even if the mediated
coupling strength is kept at a fixed level. Furthermore, it is
possible to adjust the system parameters such that the conditions
of vanishing $J$ and $J_{\rm residual}$ are simultaneously
satisfied. An example of this situation is shown in Fig.~2. Note
that the achievable effective coupling strength for the
parametric-coupling scheme has been reduced from that in Fig.~1,
because the dc bias point has now moved away from the desirable
point $\theta_{\rm C}\approx 0.65$ mentioned above.

Another relevant quantity when analyzing the ideality of the off
state is how the effective coupling strength changes in response
to changes in the coupler's bias point. This quantity can be
derived easily from Eq.~(\ref{eq:J}):
\begin{equation}
\frac{dJ}{d\epsilon_{\rm C}} = \frac{6 J_{\rm 1C} J_{\rm 2C}
\cos^4\theta_{\rm C} \sin\theta_{\rm C}}{\Delta_{\rm C}^2}.
\end{equation}
With proper design parameters, this quantity also decreases
rapidly with increasing $\Delta_{\rm C}$.

\subsection{Entanglement in the energy eigenstates}

In addition to the residual-coupling Hamiltonian, another natural
question to ask when analyzing the ideality of the off state is
how much entanglement there is between the qubits and the coupler,
and between the qubits themselves, in the different energy
eigenstates. Unlike the residual-coupling Hamiltonian, where we
had to resort to numerical calculations, this entanglement can be
evaluated using the analytic approach. The analysis of how the
energy-eigenstate entanglement manifests itself as errors in a
quantum calculation will be postponed until Sec.~V.

Here we are interested in the nearly ideal situation where any
residual entanglement is treated as a small error. We therefore
take the situation analyzed in the previous subsection
($J=\tilde{\epsilon}_1=\tilde{\epsilon}_2=0$,
$\Delta_1\neq\Delta_2$, and the coupler is a two-level system),
and we analyze the degree of entanglement to lowest non-vanishing
orders.

As explained earlier in this section, we can start our calculation
using the states $|00\underline{0}\rangle$,
$|01\underline{0}\rangle$, $|10\underline{0}\rangle$ and
$|11\underline{0}\rangle$. For the lowest order of this
calculation, we can use the pseudo-degenerate perturbation theory
calculation explained in the Appendix. We find that
\begin{eqnarray}
|\psi_{00}\rangle & \approx & |00\underline{0}\rangle -
\frac{J_0}{2\eta_1} |11\underline{0}\rangle -
\frac{A_{10}}{\eta_1} |10\underline{1}\rangle -
\frac{B_{10}}{\eta_1} |01\underline{1}\rangle \nonumber
\\
|\psi_{01}\rangle & \approx & |01\underline{0}\rangle -
\frac{J_0}{2\eta_1} |10\underline{0}\rangle -
\frac{A_{10}}{\eta_1} |11\underline{1}\rangle -
\frac{B_{10}}{\eta_1} |00\underline{1}\rangle \nonumber
\\
|\psi_{10}\rangle & \approx & |10\underline{0}\rangle -
\frac{J_0}{2\eta_1} |01\underline{0}\rangle -
\frac{A_{10}}{\eta_1} |00\underline{1}\rangle -
\frac{B_{10}}{\eta_1} |11\underline{1}\rangle \nonumber
\\
|\psi_{11}\rangle & \approx & |11\underline{0}\rangle -
\frac{J_0}{2\eta_1} |00\underline{0}\rangle -
\frac{A_{10}}{\eta_1} |01\underline{1}\rangle -
\frac{B_{10}}{\eta_1} |10\underline{1}\rangle,
\end{eqnarray}
up to a normalization constant slightly smaller than 1. Note that
the above expressions can be collectively summarized as
\cite{EigenstateCorrections}:
\begin{widetext}
\begin{eqnarray}
|\psi_{nm}\rangle & \approx & |nm\underline{0}\rangle -
\frac{J_0}{2\eta_1} |\bar{n}\bar{m}\underline{0}\rangle -
\frac{A_{10}}{\eta_1} |\bar{n}m\underline{1}\rangle -
\frac{B_{10}}{\eta_1} |n\bar{m}\underline{1}\rangle \nonumber
\\
& = & \left( 1 - \frac{J_0}{2\eta_1} \hat{\sigma}_x^{(1)} \otimes
\hat{\sigma}_x^{(2)} - \frac{A_{10}}{\eta_1} \hat{\sigma}_x^{(1)}
\otimes \hat{\sigma}_x^{(C)} - \frac{B_{10}}{\eta_1}
\hat{\sigma}_x^{(2)} \otimes \hat{\sigma}_x^{(C)} \right)
|nm\underline{0}\rangle \nonumber
\\
& = & \left( 1 - \frac{J_0\cos\theta_{\rm C}}{2\Delta_{\rm C}}
\hat{\sigma}_x^{(1)} \otimes \hat{\sigma}_x^{(2)} - \frac{J_{\rm
1C}\cos^2\theta_{\rm C}}{\Delta_{\rm C}} \hat{\sigma}_x^{(1)}
\otimes \hat{\sigma}_x^{(C)} - \frac{J_{\rm 2C}\cos^2\theta_{\rm
C}}{\Delta_{\rm C}} \hat{\sigma}_x^{(2)} \otimes
\hat{\sigma}_x^{(C)} \right) |nm\underline{0}\rangle.
\label{eq:Eigenstates}
\end{eqnarray}
\end{widetext}

Two notes are in order here: (1) The mixing of computational-basis
states in the energy eigenstates (Eq.~\ref{eq:Eigenstates}) cannot
be identified as arising from a simple residual-coupling
Hamiltonian. In other words, one cannot find a two-qubit
Hamiltonian that reproduces any of the corrections in
Eq.~(\ref{eq:Eigenstates}). This can be perhaps most clearly seen
in the fact that the signs in front of the small terms in
Eq.~(\ref{eq:Eigenstates}) are independent of the state. In
contrast, one always expects different states to acquire
corrections of different signs when dealing with a direct coupling
Hamiltonian. Higher-order effects in this system therefore cannot
be completely described by simply using small corrections to the
reduced (i.e.~$4\times4$) two-qubit Hamiltonian. (2) The above
expressions were obtained assuming only that the coupling strength
$J$ vanishes, $\tilde{\epsilon}_1=\tilde{\epsilon}_2=0$ and
$\Delta_1\neq\Delta_2$. They are not affected by the value of the
residual-coupling Hamiltonian discussed in Sec.~IV.B. This fact
demonstrates that the vanishing of both $J$ and $J_{\rm residual}$
does not imply that the qubits are entirely decoupled. Corrections
to the energy eigenstates still have to be considered when
calculating possible errors, as will be explained in Sec.~V.

As in the case of the residual-coupling Hamiltonian, it is
encouraging for scalability considerations that the entanglement
in the energy eigenstates decreases and approaches zero with
increasing $\Delta_{\rm C}$, assuming that $J_0$ and $J_1$ are
kept fixed. One can therefore say that the coupler becomes more
and more ideal as $\Delta_{\rm C}$ is increased. It should be
noted that the superconducting gap for aluminum is $\sim$ 50 GHz.
It would be interesting to investigate in the future how serious a
constraint this number imposes on the maximum allowed value of
$\Delta_{\rm C}$.

\subsection{Harmonic oscillator as a coupler}

We can use the results derived above for a two-level coupler to
infer the corresponding results for a harmonic-oscillator coupler.
We now consider the situation where
\begin{eqnarray}
\hat{H}_{\rm C} & = & \omega_{\rm C} \; \hat{a}^{\dagger} \hat{a}
\nonumber
\\
\hat{H}_{\rm 1C} & = & J_{\rm 1C} \; \hat{\sigma}_x^{(1)} \otimes
\left( \hat{a} + \hat{a}^{\dagger} \right) \nonumber
\\
\hat{H}_{\rm 2C} & = & J_{\rm 2C} \; \hat{\sigma}_x^{(2)} \otimes
\left( \hat{a} + \hat{a}^{\dagger} \right),
\end{eqnarray}
where $\hat{a}^{\dagger}$ and $\hat{a}$ are the creation and
annihilation operators for the coupler. The above coupling
Hamiltonians change the state of the coupler by one excitation. It
should therefore be a good approximation to truncate the coupler
to its lowest two energy levels. We now find that
\begin{equation}
J = J_0 - \frac{2 J_{\rm 1C} J_{\rm 2C}}{\omega_{\rm C}}
\label{eq:Jharmonic_oscillator}
\end{equation}
and
\begin{equation}
J_{\rm residual} \approx 4 \frac{\Delta_1 \Delta_2 J_0 J_{\rm 1C}
J_{\rm 2C}}{\omega_{\rm C}^4}.
\end{equation}

In order to study the tunability of a harmonic-oscillator coupler,
we now consider the effect of an applied external field on the
mediated coupling, adding the term
\begin{equation}
\hat{H}_{\rm field} = h \left( \hat{a} + \hat{a}^{\dagger} \right)
\label{eq:Linear_field}
\end{equation}
to the Hamiltonian. By defining $\hat{b} \equiv \hat{a} +
2h/\omega_{\rm C}$, we can see that the new Hamiltonian takes the
same form it had in the absence of $\hat{H}_{\rm field}$, except
that $\hat{a}$ and $\hat{a}^{\dagger}$ are replaced by $\hat{b}$
and $\hat{b}^{\dagger}$, and the qubit bias parameters
$\epsilon_1$ and $\epsilon_2$ are shifted to $\tilde{\epsilon}_1$
and $\tilde{\epsilon}_2$ (just as discussed in the general case
above). In particular, the values of $J$ and $J_{\rm residual}$
are not affected by the applied field. The mediated coupling is
therefore not tunable, assuming that the coupler's bias point is
set by an applied linear field, i.e.~a field that affects the
coupler according to Eq.~(\ref{eq:Linear_field}).

If the frequency of the harmonic oscillator (or alternatively
$J_{\rm 1C}$ and $J_{\rm 2C}$) were tunable, one would be able to
obtain a tunable value of $J$ (see
Eq.~\ref{eq:Jharmonic_oscillator}). Possible designs for such
tunable couplers have been proposed theoretically
\cite{Wallquist2}, but they have not been realized experimentally.
Given the advantages they could provide in terms of tunable
coupling, it would be highly desirable to fabricate such tunable
oscillators in the future.

A related design would be to use a (possibly non-tunable)
harmonic-oscillator element, which can be made large in size, in
the circuit and add a tunable coupler between this oscillator and
each qubit around it. With this design one would gain the
advantages of both (1) the large `cavity' being able to function
as a data bus connecting a large number of qubits and (2) the
coupling being tunable. With this design one would also be able to
use the parametric-coupling scheme to perform entangling
operations on any qubit in the circuit and the cavity relatively
easily (using red- or blue-sideband transitions), with the qubits
biased at their optimal points.

\subsection{A coupler that is near resonance with the qubits}

So far, we have focused on the case where the coupler's excitation
energy is the largest energy scale in the problem. For
completeness we consider in this subsection the case where the
coupler's excitation energy is close to those of the two qubits,
but with sufficient detuning to avoid exciting the coupler. In
other words, we consider a situation similar to the experiment of
Ref.~\cite{Majer2}, but not that of Ref.~\cite{Sillanpaa}.

The largest energy scale in the problem is now the qubit and
coupler energy splittings, which are almost equal. The next
largest energy scale is the detuning between the qubits and the
coupler (this assumption is made in order to avoid large
entanglement between the qubits and the coupler in the energy
eigenstates, in which case we would have to deal with real
excitations in the coupler). The detuning between the qubits,
direct interqubit coupling strength and qubit-coupler coupling
strength can take any values.

Because the qubit energy scale is the largest energy scale in the
problem, the mediated-coupling term now describes an excitation
transfer between the two qubits. As such, the form of the
effective-coupling term will only make sense in connection with
the single-qubit Hamiltonians, not the physically defined
operators $\hat{\sigma}_x$ and $\hat{\sigma}_z$ as in the case of
a high-excitation-energy coupler. Assuming that the qubits are
biased at their optimal points and the coupler is a harmonic
oscillator with only one relevant excited state (and using the
notation of Sec.~IV.D), the mediated-coupling term for this
excitation-transfer process is well described by the Hamiltonian
\begin{equation}
\hat{H}_{\rm mediated} = - \; \frac{J_{\rm 1C} J_{\rm
2C}}{\omega_{\rm C} - \bar{\Delta}} \left(\hat{\sigma}_+^{(1)}
\hat{\sigma}_-^{(2)} + h.c. \right),
\end{equation}
where $\bar{\Delta}=(\Delta_1+\Delta_2)/2$ (note that
$|\Delta_1-\Delta_2|$ is assumed to be much smaller than
$|\omega_{\rm C}-\bar{\Delta}|$; note also that the qubit
splittings $\Delta_j$ will be slightly modified because of the
interaction with the coupler). Using numerical calculations we
find that the residual-coupling term is given by
Eq.~(\ref{eq:Hresidual}) with
\begin{equation}
J_{\rm residual} \approx \frac{J_0 J_{\rm 1C} J_{\rm 2C}}{\left(
\omega_{\rm C} - \bar{\Delta} \right)^2}.
\end{equation}
The fact that the residual-coupling strength scales inversely with
the second power of the qubit-coupler detuning as opposed to the
fourth power of the coupler's energy splitting indicates that
residual coupling can be a more serious issue in this case than in
the case of a high-excitation-energy coupler. It should be noted,
however, that in the experiment of Ref.~\cite{Majer2} $J_0$ was
essentially zero, in which case residual coupling should be
negligible.

\section{Estimating errors in a typical experiment}

In this section we discuss a typical present-day experimental
procedure. Using concrete examples, we will be able to discuss
rather clearly the possible errors that the residual coupling and
energy-eigenstate entanglement might cause.

In most, if not all, present-day experiments, the qubits are
controlled using a single microwave line. This situation, however,
cannot be maintained for larger multi-qubit systems, where the
energy levels of the entire system become densely packed. We shall
therefore assume that local control lines are used to address the
individual qubits.

\subsection{Two-qubit system}

The usual theoretical approach to describing a quantum system that
is composed of several distinct objects commonly runs along the
following line of reasoning: First we assume that these objects
are in a separable initial state, then we controllably evolve the
system using the total Hamiltonian, we take the trace over the
degrees of freedom of the unused objects (in this case the
coupler), and we find the answer to the physical questions of
interest. Given the entangled form of the energy eigenstates
obtained in Sec.~IV, one might expect that this `contamination' of
the states will reduce the observed coherence effects. Indeed,
following the above-described recipe, one finds such a reduction
in observable coherence effects, e.g.~reduced gate fidelities. As
we now explain, however, this is not the correct description of a
typical experiment, and the resulting errors and limitations are
also described incorrectly in this way.

First let us consider the assumption about the initial conditions.
In a real two-qubit experiment (assuming essentially zero
temperature), the system starts in its ground state
$|\psi_{00}\rangle \neq |00\underline{0}\rangle$. When preparing
the desired initial state (e.g.~a product of two single-qubit
states), single-qubit operations are performed using
small-amplitude pulses that are resonant with individual qubits.
These pulses are typically weak enough that the single-qubit
operations are performed over times that are long compared to
$1/|\Delta_1-\Delta_2|$ (the energy difference
$|\Delta_1-\Delta_2|$ is used here because it is, to a good
approximation, the smallest energy difference in the spectrum of
the four lowest energy levels). Rather than perform ideal
single-qubit operations, such weak pulses drive transitions
between the eigenstates of the Hamiltonian of the entire system,
regardless of the form of these eigenstates. One could therefore
ignore the fact that $|\psi_{nm}\rangle \neq
|nm\underline{0}\rangle$ and simply use the eigenstates of the
Hamiltonian (i.e.~$|\psi_{00}\rangle$, $|\psi_{01}\rangle$,
$|\psi_{10}\rangle$ and $|\psi_{11}\rangle$) as the computational
basis for performing a quantum algorithm. One can then use weak
pulses with properly calibrated frequencies to perform any
operation on this effective two-qubit system. Thus the initial and
all subsequent states in the experiment can be very close to the
desired form, except that the basis states do not have the simple,
separable form [e.g., the state
$(|\psi_{00}\rangle+|\psi_{10}\rangle)/\sqrt{2}$ is prepared
instead of the state
$(|00\underline{0}\rangle+|10\underline{0}\rangle)/\sqrt{2}$].

\begin{figure}[h]
\includegraphics[width=7.0cm]{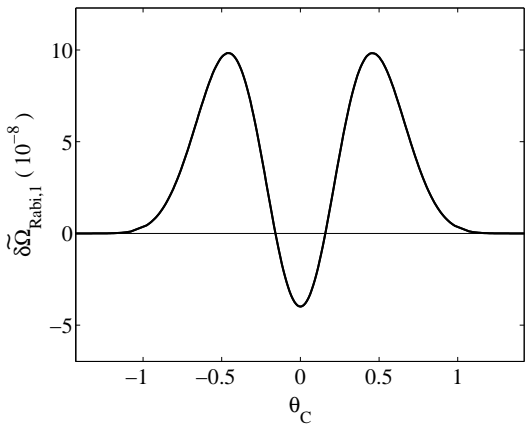}
\caption{The error-estimating quantity
$\widetilde{\delta\Omega}_{\rm Rabi,1} \equiv ( \langle \psi_{00}
| \hat{\sigma}_x^{(1)} | \psi_{10} \rangle - \langle \psi_{01} |
\hat{\sigma}_x^{(1)} | \psi_{11} \rangle ) /2$ for the relative
spread in the Rabi frequency of qubit 1 as a function of the
coupler's bias point $\theta_{\rm C}$ for the same parameters as
in Fig.~2. Similar results would be obtained for qubit 2.}
\end{figure}

There are two main possible sources of errors that can affect the
above picture. They are both related to the fact that in most
quantum algorithms one needs to perform operations on a certain
qubit without knowing the states of the other qubits. The
residual-coupling Hamiltonian (Eq.~\ref{eq:Hresidual}) provides
the first source of errors. For example, the resonance frequencies
for the transitions $|\psi_{00}\rangle \leftrightarrow
|\psi_{10}\rangle$ and $|\psi_{01}\rangle \leftrightarrow
|\psi_{11}\rangle$ are shifted from each other by a frequency
difference given by $4J_{\rm residual}$. As a result, the
resonance frequency for performing operations on qubit 1 depends
on the state of qubit 2, and vice versa. In principle, one could
perform a single-qubit gate using two pulses, with each pulse
being resonant with one of the transitions. This approach,
however, is impractical for many-qubit systems, where resonance
lines generally split into an exponentially large number of lines.
A more scalable alternative is to perform single-qubit operations
using driving amplitudes that are large compared to the spread in
the relevant resonance frequencies, such that the pulse can be
considered on resonance for all the relevant transitions. However,
the residual-coupling Hamiltonian will then cause an undesirable
phase accumulation over the course of running the algorithm. Until
this residual coupling is suppressed in future experiments
(e.g.~by increasing $\Delta_{\rm C}$ as discussed in Sec.~IV.B),
one might need to apply refocussing pulses to reduce its effects.
As can be seen from Figs.~1 and 2, pushing $J_{\rm residual}$ to
the kHz range should be possible using realistic experimental
parameters. The other possible source of errors is the fact that
the Rabi frequency of qubit 1 oscillations generally depends on
the state of qubit 2, and vice versa (This splitting of the Rabi
frequencies is a result of the mixing in the energy eigenstates).
Such errors can be described by quantities of the form
$\widetilde{\delta\Omega}_{\rm Rabi,1} \equiv ( \langle \psi_{10}
| \hat{\sigma}_x^{(1)} |\psi_{00}\rangle - \langle \psi_{11} |
\hat{\sigma}_x^{(1)} | \psi_{01} \rangle )/2$. It is
straightforward to see that, to the lowest order given in
Sec.~IV.C,
\begin{eqnarray}
\langle \psi_{10} | \hat{\sigma}_x^{(1)} |\psi_{00}\rangle & = &
\langle \psi_{11} | \hat{\sigma}_x^{(1)} |\psi_{01}\rangle.
\end{eqnarray}
In order to obtain an estimate for $\widetilde{\delta\Omega}_{\rm
Rabi,1}$, we perform some numerical calculations. We show in
Fig.~3 the results for the parameters of Fig.~2. With these
parameters we can see that $\widetilde{\delta\Omega}_{\rm Rabi,1}
\sim 10^{-8}$. This type of errors can therefore be made extremely
small using realistic experimental parameters.

A third, but less serious, point of potential concern relates to
quantities of the form $\langle \psi_{00} | \hat{\sigma}_x^{(1)}
|\psi_{01}\rangle$. This quantity describes, in some sense, how
much qubit 2 `feels' a signal applied to qubit 1 (Similarly to the
splitting of Rabi frequencies mentioned above, this type of error
is a result of the mixing in the energy eigenstates). Using the
parameters of Fig.~2, the error estimator $\langle \psi_{00} |
\hat{\sigma}_x^{(1)} |\psi_{01}\rangle$ is of order $10^{-2}$, and
vanishes close to the point where $J=0$. This type of errors can
be suppressed further by using microwave amplitudes such that the
small fraction of a signal applied to qubit 1 that is felt by
qubit 2 is small enough to be considered far off resonance.
Alternatively, one could say that single-qubit operations on qubit
1 must be performed on a time scale that is long compared to $|
\langle \psi_{00} | \hat{\sigma}_x^{(1)} |\psi_{01}\rangle | /
|\Delta_1-\Delta_2|$. This is a very mild constraint, even for
existing experiments.

Finally, a tricky issue related to the usual theoretical approach
mentioned at the beginning of this subsection is when the coupler
degrees of freedom are traced out at the end of the actual quantum
calculation. Here again, the measurement process in typical
experiments does not follow the simple picture of a sudden,
accurate measurement device that probes the basis 0/1. For flux
qubits, for example, one typically designs the readout device to
simply be a very accurate device for measuring magnetic fields
(measuring the flux generated by the qubits and going through the
loop of a readout SQUID). If the qubit-SQUID coupling is weak
compared to the qubits' energy scale and it acts for a
sufficiently long duration, the readout device will perform the
measurement in the basis of the eigenstates of the system
Hamiltonian, not the physically defined clockwise and
counterclockwise current states. If, in addition, the relevant
decoherence rates in the SQUID are small enough, the readout
fidelity will not be limited by the exact amount of mixing between
the clockwise and counterclockwise current states in the energy
eigenstates \cite{WeakMeasurement}. Therefore, the readout device
can, in principle, give the correct reading (0 or 1) essentially
100\% of the time even if the energy eigenstates that are being
used in the experiment contain finite amplitudes of the `wrong'
state [In particular, here we have in mind the states in
Eq.~(\ref{eq:Eigenstates})]. We should emphasize here that
present-day readout techniques are far from this ideal limit.

\subsection{Multi-qubit systems}

We now generalize some of the results concerning possible errors
in a typical experiment to the case where the system contains more
than two qubits. The discussion below also demonstrates an
interesting issue related to the significance of setting $J=0$ in
the off state (here we have in mind the parametric coupling
scheme).

In order to clearly identify whether a given effect is related to
mediated coupling or not, we start by considering a multi-qubit
system with no couplers (The expressions obtained below for the
errors also apply if we integrate out the couplers in the circuit
and use the effective, i.e.~direct-plus-mediated, coupling
strengths between the qubits). We first take the two-qubit
Hamiltonian:
\begin{equation}
\hat{H} = \frac{\Delta_1}{2} \hat{\sigma}_z^{(1)} +
\frac{\Delta_2}{2} \hat{\sigma}_z^{(2)} + J_{12}
\hat{\sigma}_x^{(1)} \otimes \hat{\sigma}_x^{(2)}.
\end{equation}
The above Hamiltonian has the interesting property that, with the
proper assignment of the labels $00$, $01$, $10$ and $11$ to the
eigenstates of the Hamiltonian, the relations
\begin{eqnarray}
E_{10}-E_{00} & = & E_{11}-E_{01} \nonumber
\\
E_{01}-E_{00} & = & E_{11}-E_{10} \nonumber
\\
\langle \psi_{10} | \hat{\sigma}_x^{(1)} |\psi_{00}\rangle & = &
\langle \psi_{11} | \hat{\sigma}_x^{(1)} |\psi_{01}\rangle
\nonumber
\\
\langle \psi_{01} | \hat{\sigma}_x^{(2)} |\psi_{00}\rangle & = &
\langle \psi_{11} | \hat{\sigma}_x^{(2)} |\psi_{10}\rangle
\label{eq:NiceRelations}
\end{eqnarray}
hold regardless of the values of the different parameters in the
Hamiltonian. The above relations imply that the first two sources
of error discussed in the previous subsection vanish completely
for this system. It is therefore not required to have
$J_{12}<|\Delta_1-\Delta_2|$ in order to perform an almost
error-free quantum algorithm in this system.

The question now is whether this situation persists for a system
with more than two qubits. We approach this question by performing
numerical simulations of a one-dimensional chain of three to ten
qubits with the Hamiltonian:
\begin{equation}
\hat{H} = \sum_{j=1}^N \frac{\Delta_j}{2} \hat{\sigma}_z^{(j)} +
\sum_{j=1}^{N-1} J_{j,j+1} \hat{\sigma}_x^{(j)} \otimes
\hat{\sigma}_x^{(j+1)}.
\end{equation}
From the numerical calculations we find that the generalized
version of the relations in Eq.~(\ref{eq:NiceRelations}) holds
only for the two qubits at the ends of the chain. The generalized
version of the resonance-frequency relation [i.e.~the first two
lines in Eq.~(\ref{eq:NiceRelations})] continues to hold for all
the qubits (i.e.~the resonance frequency for flipping qubit $j$ is
independent of the states of all the other qubits). The Rabi
frequency of qubit $j$ oscillations, however, now generally
depends on the states of the other qubits.

In order to get an estimate for the above-mentioned errors, we
take the three-qubit case and calculate the standard deviation
$\widetilde{\delta\Omega}_{\rm Rabi,2}$ in the quantity $\langle
\psi_{n 1 m} | \hat{\sigma}_x^{(2)} | \psi_{n 0 m} \rangle$. When
the coupling strengths $J_{j,j+1}$ are small compared to the
differences between qubit gaps, we find that
\begin{equation}
\widetilde{\delta\Omega}_{\rm Rabi,2} \propto
\frac{J_{12}^2J_{23}^2}{\left| (\Delta_1-\Delta_2)
(\Delta_2-\Delta_3) (\Delta_1-\Delta_3)^2 \right|},
\end{equation}
with a prefactor of order one (we also find that this expression
remains valid when we add a few more qubits at the ends of the
chain). It is for this reason that one would like to set $J=0$
(for all pairs of qubits) in the off state, even though violating
this requirement does not necessarily have detrimental effects on
a two-qubit experiment \cite{Rigetti}. It will also be highly
desirable to set the gaps $\Delta_j$ to different values, even if
they are not nearest neighbours.

We now take a three-qubit chain with couplers, and we perform
numerical simulations for several sets of parameters with the
leading-order coupling strengths (i.e.~the parameters that
correspond to $J$ in the two-qubit case) set to zero. We calculate
the resonance frequencies for qubit $j$ ($j=1,2,3$) for the four
different states of the two other qubits. We find that the
resonance-frequency shifts are consistent with the residual
coupling strength given in Eq.~(\ref{eq:ResidualJ}). We therefore
expect our results concerning the residual coupling to hold for
systems with more than two qubits. We also expect the form of the
energy eigenstates to follow the rather straightforward
generalization of Eq.~(\ref{eq:Eigenstates}). If a distant pair of
qubits have the same value of $\Delta_j$, the energy scale
associated with any possible hybridization of energy levels
(e.g.~involving states of the form $|0 \underline{0} n_2
\underline{0} n_3 \underline{0} n_4 \underline{0} 1 \underline{0}
n_6 ...\rangle$ and $|1 \underline{0} n_2 \underline{0} n_3
\underline{0} n_4 \underline{0} 0 \underline{0} n_6 ...\rangle$)
will be small enough that it can be neglected. Similarly,
hybridization between energy levels that require flipping the
states of several qubits should be greatly suppressed.

Next we consider a loop of three or four qubits, i.e.~in triangle
and square geometries. Denoting the typical scale of the coupling
strengths between neighbouring qubits by $J$ and the detuning
between the qubits by $\Delta\omega$, which is taken to be much
smaller than the qubit frequencies, we find that the standard
deviation in the resonance frequency $\delta \omega_{\rm
resonance}$ and the standard deviation in the transition matrix
element $\widetilde{\delta\Omega}_{\rm Rabi}$ when attempting to
change the state of a given qubit are given by
\begin{eqnarray}
\delta\omega_{\rm resonance} \sim \frac{J^3}{\Delta\omega^2}
\nonumber
\\
\widetilde{\delta\Omega}_{\rm Rabi} \sim \left(
\frac{J}{\Delta\omega} \right)^3
\end{eqnarray}
in the three qubit case and
\begin{eqnarray}
\delta\omega_{\rm resonance} \sim \frac{J^4}{\Delta\omega^3}
\nonumber
\\
\widetilde{\delta\Omega}_{\rm Rabi} \sim \left(
\frac{J}{\Delta\omega} \right)^4
\end{eqnarray}
in the four qubit case. The detailed expressions for the above
quantities depend rather nontrivially on the values of the
different parameters in the problem. Errors stemming from the
spreads in resonance and Rabi frequencies therefore depend on the
geometry of the multi-qubit system, e.g.~chain versus closed loop.
A two-dimensional lattice would belong to the latter category.

Finally, we consider the geometry where every qubit is coupled to
every other qubit, which is relevant to the case of a single
coupler mediating coupling between several qubits. We find that
the scaling of errors in this case follows the triangle geometry
discussed above, with $\Delta\omega$ representing the typical
spacing of frequencies in the system. In particular, adding qubits
to the system at distant frequencies has little effect on the
errors associated with a given qubit.

The main result of this subsection is the fact that certain errors
that have not been relevant to past experiments (i.e.~two-qubit
experiments) will likely arise in future experiments. Although
this result might seem somewhat discouraging, it is encouraging
that the errors, quantified by the spreads in the resonance and
Rabi frequencies, follow scaling laws that make them less serious
than one might intuitively expect
(e.g.~$\widetilde{\delta\Omega}_{\rm Rabi}$ scales as the third or
fourth power of $J/\Delta\omega$, as opposed to being linearly
proportional to $J/\Delta\omega$).

\section{conclusion}

We have analyzed the problem of a high-excitation-energy quantum
object mediating coupling between two qubits. After reviewing some
known results concerning the leading-order term in the mediated
coupling, we obtained expressions that describe the
residual-coupling in the off state and the entanglement in the
energy eigenstates of the system. We have argued that our approach
analyzing the properties of the eigenvalues and eigenstates of the
total Hamiltonian is the appropriate one to describe recent, and
possibly future, experiments. We have also estimated the expected
errors originating from the non-ideality of the off state in
typical experimental situations. Our results should be helpful in
designing future circuits that employ the mediated-coupling
approach in order to achieve tunable coupling between qubits. In
particular, our results suggest that with properly chosen design
parameters, the residual coupling in the off state could be
greatly reduced in future experiments.

We have focused on the case of a two-qubit system, which is the
relevant case for present-day experiments. However, we expect our
results for the errors in a two-qubit system to apply to
multi-qubit systems as well. As we have shown, other sources of
error that are absent in a two-qubit system arise for systems with
more than two qubits. It would therefore be interesting and
important in the future to analyze in more detail the properties
of large, many-qubit systems.

Finally, we should mention that although a large part of our
analysis was formulated in the language of superconducting
systems, our results are quite general and should apply to other
systems that employ similar mediated-coupling mechanisms
\cite{Hodges}.

\appendix

\begin{center}
{\bf Appendix: Quasi-degenerate perturbation theory}
\end{center}

In this appendix we present a perturbation-theory procedure for
the case where the separation between some energy levels is not
large compared to the energy scale of the perturbation
\cite{Sakurai}.

Let us take the Hamiltonian
\begin{equation}
\hat{H} = \hat{H}_0 + \hat{V},
\end{equation}
where we want to treat $\hat{V}$ as a perturbation. We assume,
however, that the energy scale of $\hat{V}$ is not necessarily
small compared to the energy separations within a subset of $n$
eigenstates of $\hat{H}_0$. The energy scale of $\hat{V}$ is small
compared to the energy separation between this subset of levels
and all the levels outside it.

We now want to find the approximate energy levels and eigenstates
of $\hat{H}$ in the vicinity of these $n$ original energy levels
of $\hat{H}_0$. We proceed by assuming that these closely spaced
eigenstates can have a large amount of mixing among themselves
because of the added perturbation $\hat{V}$, but we assume that
mixing with all other states will be small. In other words, we
express (any one of) the eigenstates of interest as
\begin{equation}
|\Psi_i\rangle = \sum_{j=1}^{n} f_{ij} |\psi_j\rangle +
\sum_{j=n+1}^{\infty} g_{ij} |\psi_j\rangle
\end{equation}
where $|\psi_j\rangle$ are the eigenstates of $\hat{H}_0$ with
\begin{equation}
\hat{H}_0 |\psi_j\rangle = \epsilon_j |\psi_j\rangle,
\end{equation}
the states $|\Psi_i\rangle$ with $j=1,2,...,n$ represent the
states of interest, and $g_{ij}$ are understood to be small enough
to be treated perturbatively (this is the only reason we express
the amplitudes $f_{ij}$ and $g_{ij}$ using two different symbols).
We can now express the eigenvalue problem as
\begin{widetext}
\begin{equation}
\sum_{j=1}^{n} f_{ij} \epsilon_j |\psi_j\rangle +
\sum_{j=n+1}^{\infty} g_{ij}  \epsilon_j |\psi_j\rangle +
\sum_{k=1}^{\infty} \left( \sum_{j=1}^{n} f_{ij} V_{kj}
|\psi_k\rangle + \sum_{j=n+1}^{\infty} g_{ij} V_{kj}
|\psi_k\rangle \right) = E_i \left( \sum_{j=1}^{n} f_{ij}
|\psi_j\rangle + \sum_{j=n+1}^{\infty} g_{ij} |\psi_j\rangle
\right).
\end{equation}
\end{widetext}
If we multiply the above equation from the left by $\langle
\psi_l|$ with $1\leq l \leq n$, we obtain the equation
\begin{equation}
f_{il} \epsilon_l + \sum_{j=1}^{n} f_{ij} V_{lj} +
\sum_{j=n+1}^{\infty} g_{ij} V_{lj} = E_i f_{il}.
\label{eq:PerturbSchEq1}
\end{equation}
If we multiply the equation by $\langle \psi_l|$ with $l>n$, we
obtain instead:
\begin{equation}
g_{il} \epsilon_l + \sum_{j=1}^{n} f_{ij} V_{lj} +
\sum_{j=n+1}^{\infty} g_{ij} V_{lj} = E_i \; g_{il}
\end{equation}
which can be rewritten as
\begin{equation}
g_{ij} = \frac{ \sum_{j=1}^{n} f_{ij} V_{lj} +
\sum_{j=n+1}^{\infty} g_{ij} V_{lj} }{ E_i - \epsilon_l }
\label{eq:PerturbSchEq2}
\end{equation}
Equation (\ref{eq:PerturbSchEq2}) is now treated using a
perturbative (or rather iterative) approach. If we neglect the sum
over the states with $j=n+1,...$, we obtain:
\begin{equation}
g_{il} \approx \frac{\sum_{j=1}^{n} f_{ij}
V_{lj}}{E_i-\epsilon_l}.
\label{eq:FirstIteration}
\end{equation}
Substituting this expression into Eq.~(\ref{eq:PerturbSchEq1}), we
obtain (for every $l$ with $1 \leq l \leq n$):
\begin{equation}
\sum_{j=1}^{n} \left( \epsilon_l \delta_{lj} + V_{lj} +
\sum_{j'=n+1}^{\infty} \frac{ V_{lj'} V_{j'j}}{E_i-\epsilon_{j'}}
\right) f_{ij} = E_i f_{il},
\label{eq:1stOrderSchEq}
\end{equation}
where $\delta_{lj}$ is the Kronecker delta function.

If we want to obtain more accurate results, we can take
Eq.~(\ref{eq:FirstIteration}) and substitute it into the
right-hand side of Eq.~(\ref{eq:PerturbSchEq2}). Expressing
$g_{ij}$ in Eq.~(\ref{eq:FirstIteration}) as $g_{ij}^{\rm prev}$,
we obtain:
\begin{widetext}
\begin{eqnarray}
g_{il} & \approx & \frac{\sum_{j=1}^{n} f_{ij}
V_{lj}}{E_i-\epsilon_l} + \frac{1}{E_i-\epsilon_l} \left(
\sum_{j=n+1}^{\infty} g_{ij}^{\rm prev} V_{lj} \right) \nonumber
\\
& = & \frac{\sum_{j=1}^{n} f_{ij} V_{lj}}{E_i-\epsilon_l} +
\frac{1}{E_i-\epsilon_l} \left( \sum_{j=n+1}^{\infty}
\frac{\sum_{j'=1}^{n} f_{ij'} V_{jj'}}{E_i-\epsilon_j} V_{lj}
\right).
\end{eqnarray}
Using this expression for $g_{ij}$ in
Eq.~(\ref{eq:PerturbSchEq1}), we find that to the next order:
\begin{equation}
\sum_{j=1}^{n} \left( \epsilon_l \delta_{lj} + V_{lj} +
\sum_{j'=n+1}^{\infty} \frac{ V_{lj'} V_{j'j}}{E_i-\epsilon_j'} +
\sum_{j'=n+1}^{\infty} \sum_{j''=n+1}^{\infty} \frac{V_{lj'}
V_{j'j''} V_{j''j}}{(E_i-\epsilon_{j'}) (E_i-\epsilon_{j''})}
\right) f_{ij} = E_i f_{il}.
\label{eq:3rdOrderSchEq}
\end{equation}
\end{widetext}
The generalization to all orders is now obvious, if needed. One
must be careful, of course, that the denominators on the left-hand
side of Eqs.~(\ref{eq:1stOrderSchEq}) and (\ref{eq:3rdOrderSchEq})
contain the eigenvalue $E_i$. The solution is therefore not yet
obtained by a straightforward diagonalization of an $n \times n$
matrix. However, it is usually a good first approximation to use
some (averaged) value for $E_i$ on the left-hand side of the
equation. This value can be taken from the energy levels of the
original Hamiltonian $\hat{H}_0$. One can then obtain more
accurate results by introducing a second iterative procedure.
Every time we obtain an approximate value of the energy $E_i$, we
can substitute it into the left-hand side of
Eq.~(\ref{eq:1stOrderSchEq}) or (\ref{eq:3rdOrderSchEq}) to obtain
an even more accurate result. Note that this iterative procedure
is independent of the other one introduced above, i.e.~the
iterative substitution of Eq.~(\ref{eq:FirstIteration}) into
Eq.~(\ref{eq:PerturbSchEq2}). In order to reach a given level of
accuracy, both procedures must be performed to the appropriate
order.

\begin{acknowledgments}
We would like to thank M.~Grajcar, J.~R.~Johansson, A.~Maassen van
den Brink, D.~Tsomokos and J.~Q.~You for useful discussions. This
work was supported in part by the National Security Agency (NSA),
the Laboratory for Physical Sciences (LPS), the Army Research
Office (ARO), the National Science Foundation (NSF) grant
No.~EIA-0130383 and the Japan Society for the Promotion of Science
Core-To-Core (JSPS CTC) program. One of us (S.A.) was supported by
the Japan Society for the Promotion of Science (JSPS).
\end{acknowledgments}


\begin{thebibliography}{99}

\bibitem{YouReview} For recent reviews on the subject, see e.g.
J. Q. You and F. Nori, Phys. Today {\bf 58} (11), 42 (2005); G.
Wendin and V. Shumeiko, in {\it Handbook of Theoretical and
Computational Nanotechnology}, ed. M. Rieth and W. Schommers (ASP,
Los Angeles, 2006).

\bibitem{Pashkin} Yu. A. Pashkin, T. Yamamoto, O. Astafiev, Y. Nakamura,
D. V. Averin, and J. S. Tsai, {\it Nature} {\bf 421}, 823 (2003).

\bibitem{Johnson} P. R. Johnson, F. W. Strauch, A. J. Dragt, R. C.
Ramos, C. J. Lobb, J. R. Anderson, and F. C. Wellstood, Phys. Rev.
B {\bf 67}, 020509(R) (2003).

\bibitem{Berkley} A. J. Berkley, H. Xu, R. C. Ramos, M. A. Gubrud,
F. W. Strauch, P. R. Johnson, J. R. Anderson, A. J. Dragt, C. J.
Lobb, and F. C. Wellstood, Science {\bf 300}, 1548 (2003).

\bibitem{Yamamoto} T. Yamamoto, Yu. A. Pashkin, O. Astafiev, Y.
Nakamura, and J. S. Tsai, {\it Nature} {\bf 425}, 941 (2003).

\bibitem{Izmalkov} A. Izmalkov, M. Grajcar, E. Il'ichev, Th. Wagner,
H.-G. Meyer, A. Yu. Smirnov, M. H. S. Amin, A. Maassen van den
Brink, and A. M. Zagoskin, Phys. Rev. Lett. 93, 037003 (2004).

\bibitem{Xu} H. Xu, F. W. Strauch, S. K. Dutta, P. R.
Johnson, R. C. Ramos, A. J. Berkley, H. Paik, J. R. Anderson, A.
J. Dragt, C. J. Lobb, and F. C. Wellstood, Phys. Rev. Lett. {\bf
94}, 027003 (2005).

\bibitem{McDermott} R. McDermott, R. W. Simmonds, M. Steffen,
K. B. Cooper, K. Cicak, K. D. Osborn, S. Oh, D. P. Pappas, and J.
Martinis, Science {\bf 307}, 1299 (2005).

\bibitem{Majer1} J. B. Majer, F. G. Paauw, A. C. J. ter Haar, C. J. P.
M. Harmans, and J. E. Mooij, Phys. Rev. Lett. {\bf 94}, 090501
(2005).

\bibitem{PlourdeExp} B. L. T. Plourde, T. L. Robertson, P. A. Reichardt,
T. Hime, S. Linzen, C.-E. Wu, and J. Clarke, Phys. Rev. B {\bf
72}, 060506(R) (2005).

\bibitem{GrajcarExp} M. Grajcar, A. Izmalkov, S. H. W. van der Ploeg, S.
Linzen, T. Plecenik, Th. Wagner, U. H\"ubner, E. Il'ichev, H.-G.
Meyer, A. Yu. Smirnov, P. J. Love, A. Maassen van den Brink, M. H.
S. Amin, S. Uchaikin, and A. M. Zagoskin, Phys. Rev. Lett. {\bf
96}, 047006 (2006).

\bibitem{Steffen} M. Steffen, M. Ansmann, R. C. Bialczak, N. Katz, E.
Lucero, R. McDermott, M. Neeley, E. M. Weig, A. N. Cleland, and J.
M. Martinis, Science, {\bf 313}, 1423 (2006).

\bibitem{vdPloeg} S. H. W. van der Ploeg, A. Izmalkov, A. Maassen van
den Brink, U. Huebner, M. Grajcar, E. Il'ichev, H.-G. Meyer, and
A. M. Zagoskin, Phys. Rev. Lett. {\bf 98}, 057004 (2007).

\bibitem{Harris} R. Harris, A. J. Berkley, M. W. Johnson, P. Bunyk,
S. Govorkov, M. C. Thom, S. Uchaikin, A. B. Wilson, J. Chung, E.
Holtham, J. D. Biamonte, A. Yu. Smirnov, M. H. S. Amin, and A.
Maassen van den Brink, Phys. Rev. Lett. {\bf 98}, 177001 (2007).

\bibitem{Hime} T. Hime, P. A. Reichardt, B. L. T. Plourde, T. L.
Robertson, C.-E. Wu, A. V. Ustinov, and J. Clarke , Science {\bf
314}, 1427 (2006).

\bibitem{NiskanenExp1} A. O. Niskanen, K. Harrabi, F. Yoshihara,
Y. Nakamura, and J. S. Tsai, Phys. Rev. B {\bf 74}, 220503(R)
(2006).

\bibitem{NiskanenExp2} A. O. Niskanen, K. Harrabi, F. Yoshihara, Y.
Nakamura, S. Lloyd, and J. S. Tsai, Science {\bf 316}, 723 (2007).

\bibitem{Plantenberg} J. Plantenberg, P. C. de Groot, C. J. P. M.
Harmans, and J. E. Mooij, Nature {\bf 447}, 836 (2007).

\bibitem{Sillanpaa} M. A. Sillanp\"a\"a, J. I. Park, and R. W. Simmonds,
Nature {\bf 449}, 438 (2007).

\bibitem{Majer2} J. Majer, J. M. Chow, J. M. Gambetta, J. Koch, B. R.
Johnson, J. A. Schreier, L. Frunzio, D. I. Schuster, A. A. Houck,
A. Wallraff, A. Blais, M. H. Devoret, S. M. Girvin, R. J.
Schoelkopf, Nature {\bf 449}, 443 (2007).

\bibitem{Averin} D. V. Averin and C. Bruder, Phys. Rev. Lett. {\bf 91},
057003 (2003).

\bibitem{PlourdeTh} B. L. T. Plourde, J. Zhang, K. B. Whaley, F. K.
Wilhelm, T. L. Robertson, T. Hime, S. Linzen, P. A. Reichardt,
C.-E. Wu, and J. Clarke, Phys. Rev. B {\bf 70}, 140501(R) (2004).

\bibitem{Wallquist1} M. Wallquist, J. Lantz, V. S. Shumeiko, and G.
Wendin, New J. Phys. {\bf 7}, 178 (2005).

\bibitem{MaassenvdBrink} A. Maassen van den Brink, A. J. Berkley, and M.
Yalowsky, New J. Phys. {\bf 7}, 230 (2005).

\bibitem{Makhlin} Before Refs.~\cite{Averin,PlourdeTh,MaassenvdBrink}
there had been proposals to couple superconducting charge qubits
using a large inductance; see, e.g., A. Shnirman, G. Sch\"on, and
Z. Hermon, Phys. Rev. Lett. {\bf 79}, 2371 (1997); Y. Makhlin, G.
Sch\"on, and A. Shnirman, Nature {\bf 398}, 305 (1999). Smaller,
more realistic, inductances were considered in J. Q. You, J. S.
Tsai, and F. Nori, Phys. Rev. Lett. {\bf 89}, 197902 (2002). A
similar approach for flux qubits was proposed in J. Q. You, Y.
Nakamura, and F. Nori, Phys. Rev. B {\bf 71}, 024532 (2005).

\bibitem{Kim} See also M. D. Kim and J. Hong, Phys. Rev. B {\bf
70}, 184525 (2004); M. D. Kim, Phys. Rev. B {\bf 74}, 184501
(2006).

\bibitem{Farhi} E. Farhi and S. Gutmann, Phys. Rev. A {\bf 57},
2403 (1998).

\bibitem{Bertet} P. Bertet, C. J. P. M. Harmans, and J. E. Mooij,
Phys. Rev. B. {\bf 73}, 064512 (2006).

\bibitem{NiskanenTh} A. O. Niskanen, Y. Nakamura, and J. S. Tsai,
Phys. Rev. B {\bf 73}, 094506 (2006).

\bibitem{GrajcarTh} M. Grajcar, Y.-X. Liu, F. Nori, and A. M.
Zagoskin, Phys. Rev. B {\bf 74}, 172505 (2006).

\bibitem{Liu} Performing two-qubit gates on superconducting qubits
by driving the qubits at the sum or difference of their
characteristic frequencies was first proposed in Y.-X. Liu, L.-F.
Wei, J. S. Tsai, and F. Nori, Phys. Rev. Lett. {\bf 96}, 067003
(2006).

\bibitem{Chiorescu} Experiments on a single qubit coupled to a
`cavity' were first reported in I. Chiorescu, P. Bertet, K. Semba,
Y. Nakamura, C. J. P. M. Harmans, and J. E. Mooij, Nature {\bf
431}, 159 (2004); A. Wallraff, D. I. Schuster, A. Blais, L.
Frunzio, R. S. Huang, J. Majer, S. Kumar, S. M. Girvin, and R. J.
Schoelkopf, Nature {\bf 431}, 162 (2004); See also J. Johansson,
S. Saito, T. Meno, H. Nakano, M. Ueda, K. Semba, and H.
Takayanagi, Phys. Rev. Lett. {\bf 96}, 127006 (2006).

\bibitem{CavityTheory} There are a large number of theoretical
studies on this subject. See, e.g., J. Q. You and F. Nori, Phys.
Rev. B {\bf 68}, 064509 (2003); R. Migliore, A. Konstadopoulou, A.
Vourdas, T. P. Spiller, and A. Messina, Phys. Lett. A {\bf 319},
67 (2003); Y.-X. Liu, L. F. Wei, J. R. Johansson, J. S. Tsai, and
F. Nori, Phys. Rev. B {\bf 76}, 144518 (2007); A. Blais, J.
Gambetta, A. Wallraff, D. I. Schuster, S. M. Girvin, M. H.
Devoret, and R. J. Schoelkopf, Phys. Rev. A {\bf 75}, 032329
(2007); H. Nakano, K. Kakuyanagi, M. Ueda, and K. Semba, Appl.
Phys. Lett. {\bf 91}, 032501 (2007); F. Helmer, M. Mariantoni, A.
G. Fowler, J. von Delft, E. Solano, F. Marquardt, arXiv:0706.3625.

\bibitem{Hutter} C. Hutter, A. Shnirman, Y. Makhlin, and G.
Sch\"on, Europhys. Lett. {\bf 74}, 1088 (2006).

\bibitem{Baym} See, e.g., G. Baym, {\it Lectures on Quantum
Mechanics} (Addison-Wesley, New York, 1990).

\bibitem{NoCJ} We shall not derive the expressions for $c_j$ in this
section; the relevant expressions agree, to lowest order, with
those that will be derived in Sec.~IV.

\bibitem{Sakurai} See, e.g., J. J. Sakurai, {\it Modern Quantum
Mechanics} (Addison-Wesley, New York, 1994).

\bibitem{A00B00} In the theoretical treatment of this system, one can
also rewrite the Hamiltonian such that the matrix elements
$A_{00}$ and $B_{00}$ are absorbed into $\epsilon_1$ and
$\epsilon_2$. One then replaces $A_{ll}$ by $A_{ll}-A_{00}$, and
similarly for $\hat{B}$.

\bibitem{EigenstateCorrections} The second term on the right-hand
side of Eq.~(\ref{eq:Eigenstates}) is obtained by including the
gaps $\Delta_j$ in the denominator in
Eq.~(\ref{eq:2ndOrderCorrection}), which slightly modifies $J_1$
depending on the state under consideration.

\bibitem{Wallquist2} M. Wallquist, V. S. Shumeiko, and G. Wendin,
Phys. Rev. B {\bf 74}, 224506 (2006). Note that, strictly
speaking, the coupler is no longer an harmonic, i.e.~linear,
oscillator in this design.

\bibitem{WeakMeasurement} For example, it is possible, in principle,
for a device that couples to the operator $\hat{\sigma}_z$ of a
two-level system to distinguish with essentially 100\% fidelity
between energy eigenstates of the form $\sqrt{0.6} |0\rangle +
\sqrt{0.4} |1\rangle$ and $\sqrt{0.4} |0\rangle - \sqrt{0.6}
|1\rangle$.

\bibitem{Rigetti} For coupling proposals that rely on the relation
$J<|\Delta_1-\Delta_2|$ in order to achieve effective decoupling,
see e.g. C. Rigetti, A. Blais, and M. Devoret, Phys. Rev. Lett.
{\bf 94}, 240502 (2005); G. S. Paraoanu, Phys. Rev. B {\bf 74},
140504(R) (2006); S. Ashhab, S. Matsuo, N. Hatakenaka, and F.
Nori, Phys. Rev. B {\bf 74}, 184504 (2006); S. Ashhab and F. Nori,
Phys. Rev. B {\bf 76}, 132513 (2007); See also Ref.~\cite{Liu}.

\bibitem{Hodges} For example, there has been a recent proposal to
use the high-frequency electrons as mediators for performing gates
on nuclear spins in NMR quantum computing; J. S. Hodges, J. C.
Yang, C. Ramanathan, and D. G. Cory, arXiv:0707.2956v1.

\end{thebibliography}
\end{document}